\documentclass[prl,twocolumn,nofootinbib, preprintnumbers, superscriptaddress]{revtex4-1}

\usepackage{amsmath,amssymb,slashed,braket}
\usepackage{lipsum} 
\usepackage{graphicx}
\usepackage{epstopdf}
\usepackage{float,appendix}
\usepackage[colorlinks=true,
            linkcolor=blue,
            urlcolor=blue,
            citecolor=green,          
            bookmarks=true,
            bookmarksnumbered=true,
            breaklinks=true,
            pdfpagemode=FullScreen,
            pdfstartview=FitBH]{hyperref}

\usepackage{esint}

\usepackage[normalem]{ulem}
\usepackage{color}

\definecolor{Orange}{cmyk}{0,0.61,0.87,0}
\definecolor{JungleGreen}{cmyk}{0.99,0,0.52,0}
\definecolor{OliveGreen}{cmyk}{0.64,0,0.95,0.40}
\definecolor{Brown}{cmyk}{0,0.81,1,0.60}
\definecolor{RoyalBlue}{cmyk}{0.71,0.53,0,0.12}
\definecolor{Gray}{cmyk}{0,0,0,0.40}
\definecolor{LightPink}{cmyk}{0.0,0.25,0,0}
\definecolor{LLightPink}{cmyk}{0.0,0.10,0,0}
\definecolor{LightBlue}{cmyk}{0.25,0,0,0}
\definecolor{LightGray}{cmyk}{0,0,0,0.2}

\usepackage{xcolor}
\definecolor{gesfpurple}{rgb}{0.47,0.19,0.42}

\definecolor{gesflanse}{rgb}{0.00,0.50,0.50}

\definecolor{gesfblue}{rgb}{0.08,0.42,0.76}

\definecolor{gesfred}{rgb}{1,0,0}

\definecolor{gesfwhite}{rgb}{1,1,1}

\definecolor{gesfblack}{rgb}{0,0,0}

\newcommand{\geqn}[1]{Eq.\,\hypersetup{linkcolor=blue}(\ref{#1})\hypersetup{linkcolor=blue}}
\newcommand{\gfig}[1]{{\hypersetup{linkcolor=violet}Fig.\,\ref{#1}\hypersetup{linkcolor=blue}}}

\graphicspath{{figs/}}

\begin{document}

\title{Detecting the Coupling of Axion Dark Matter to Neutron Spins at Spallation Sources via Rabi Oscillation}

\author{Peter Fierlinger}
\email{peter.fierlinger@tum.de}
\affiliation{TUM School of Natural Sciences, Technical University of Munich, 85748 Garching, Germany}

\author{Jie Sheng}
\email{shengjie04@sjtu.edu.cn}
\affiliation{Tsung-Dao Lee Institute \& School of Physics and Astronomy, Shanghai Jiao Tong University, China}
\affiliation{Key Laboratory for Particle Astrophysics and Cosmology (MOE) \& Shanghai Key Laboratory for Particle Physics and Cosmology, Shanghai Jiao Tong University, Shanghai 200240, China}

\author{Yevgeny V. Stadnik}
\email{yevgenystadnik@gmail.com}
\affiliation{School of Physics, The University of Sydney, Sydney, New South Wales 2006, Australia}

\author{Chuan-Yang Xing}
\email{cyxing@upc.edu.cn}
\affiliation{College of Science, China University of Petroleum (East China), Qingdao 266580, China}

\begin{abstract}

We propose a novel detection method for axion dark matter using the Rabi oscillation of  neutron spins in beam-based measurements. 
If axions couple to neutron spins, a background oscillating axion dark matter field would drive transitions between spin-up and spin-down neutron states in a magnetic field when the axion particle energy matches the energy gap between the spin states.
The transition can be detected in a double-Stern-Gerlach-type apparatus, with the first splitter producing a pure spin-polarized neutron beam and the second splitter selecting spin-flipped signals. 
Our approach offers enhanced detection capability for axions within the $10^{-12} - 10^{-10}$\,eV mass window with the capability to surpass the sensitivity of current laboratory experiments. 
\end{abstract}

\maketitle 

\textbf{Introduction} -- 
Dark matter (DM), comprising approximately $80\%$ of the Universe's matter content~\cite{Planck:2018vyg}, is supported by numerous astronomical and cosmological observations~\cite{Bertone:2004pz, Young:2016ala, Arbey:2021gdg, Billard:2021uyg}.
However, our understanding of the nature of DM remains limited.
Current models about DM span a wide mass spectrum, 
ranging from ultralight wave-like DM~\cite{Hui:2021tkt} to particle-like DM~\cite{Profumo:2019ujg} to macroscopic objects like primordial black holes~\cite{Carr:2020xqk, Carr:2021bzv}.
Among them, the axion has been a well-motivated DM candidate for a long time~\cite{Preskill:1982cy,Abbott:1982af,Dine:1982ah,Hui:2021tkt,Adams:2022pbo}.
The axion was originally proposed to solve the strong CP problem~\cite{Peccei:1977hh,Weinberg:1977ma,Wilczek:1977pj,Peccei:2006as}. 
String theory also predicts the existence of axion-like particles~\cite{Svrcek:2006yi}.

Depending on theoretical models, axion DM can interact with various Standard Model particles, including photons, gluons, electrons, protons, and neutrons.
Currently, most ground-based direct-detection experiments focus on detecting the axion-photon coupling.
Helioscope experiments~\cite{Sikivie:1983ip,CAST:2024eil,IAXO:2019mpb} search for axions produced in the Sun and are sensitive to a wide range of axion masses below the keV scale.
Haloscope experiments~\cite{Sikivie:1983ip,CAPP:2024dtx,QUAX:2024fut,Quiskamp:2024oet,ADMX:2024xbv}, on the other hand, concentrate on detecting axion DM in the Galactic halo and exhibit exceptional sensitivity to axions with Compton wavelengths of $\sim 1 \, \mathrm{m}$. 

In contrast, terrestrial experiments targeting axion couplings to fermions, particularly the pure coupling to neutrons, are relatively scarce. 
The most sensitive current searches for axion-nucleon couplings utilize magnetometry-based schemes 
to measure the anomalous magnetic-field-like effects of the axion DM field on nuclear or neutron spins~\cite{Abel:2017rtm,Wu:2019exd,Garcon:2019inh,Bloch:2019lcy,Jiang:2021dby,Bloch:2021vnn,Lee:2022vvb,Bloch:2022kjm,Abel:2022vfg,Wei:2023rzs,Xu:2023vfn,Gavilan-Martin:2024nlo}. 
However, these schemes mainly apply to axion DM with lighter masses $\lesssim 10^{-12}$\,eV.
For higher masses, different detection approaches are needed. 
One such potentially promising direction is the proposed CASPEr experiment~\cite{Budker:2013hfa,JacksonKimball:2017elr}.

In this paper, we propose a novel detection method for axion DM using the Rabi oscillation of neutron spins in beam-based measurements. 
In a uniform magnetic field, neutrons form a two-level system involving their spin-up and spin-down states.
Due to the axion-neutron coupling, the axion DM background field acts as an oscillating perturbation on the neutron spins with the oscillation frequency set by the energy of the axion DM particles, which owing to the non-relativistic nature of the Galactic DM is approximately given by the axion rest mass energy. 
As a result, neutron spins undergo transitions between the spin-up and spin-down states (spin-flips), with the maximum effect occurring when the axion mass-energy matches the energy level difference in the two-level system. 
This phenomenon is known as Rabi oscillation. 
We propose an experimental setup employing a double-Stern-Gerlach-type apparatus to precisely measure axion-DM-induced spin-flips in high-intensity neutron beams. 
New major neutron sources, such as the European Spallation Source (ESS)~\cite{Garoby:2017vew}, the Spallation Neutron Source (SNS) in the U.S.~\cite{MASON:2006955}, and the China Spallation Neutron Source (CSNS)~\cite{Wang_Sheng:2009} can provide strong neutron beams and facilitate the axion DM search. Ref.~\cite{Fierlinger:2024aik}
recently proposed the use of high-intensity neutron beams to probe the axion-neutron coupling using the Ramsey method; however, that method is mainly limited to rather small axion DM masses $\lesssim 10^{-14}$\,eV. 
In contrast, the Rabi oscillation approach we propose in the present work enables searches for axion DM with masses as high as $10^{-10}$\,eV.

\textbf{Rabi Oscillation of Neutron Spins in Axion Background} -- Rabi oscillation refers to the transition between two states of a two-level system under an oscillatory perturbation.
The two-level system can be constructed by applying a uniform magnetic field $\mathbf{B}_0$ to neutron beams, as
neutron spins couple to the magnetic field $\mathbf{B}_0$ via the Hamiltonian,
\begin{equation}
\label{magnetic_interaction_Hamiltonian}
    H_0 = - g_n \frac{e}{2m_p} \mathbf{S} \cdot \mathbf{B}_0 \, .
\end{equation}
Here,
$g_n = -3.826$ is the $g$-factor of the neutron, $m_p$ is the proton mass, and $\mathbf{S}$ is defined in terms of the vector of Pauli spin matrix operators $\boldsymbol{\sigma}$ according to $\mathbf{S} = \boldsymbol{\sigma}/2$. 
The energy eigenstates of this two-level system are the spin-up $\ket{\uparrow}$ and spin-down $\ket{\downarrow}$ states of neutrons
along the direction of the magnetic field, with eigenenergies $\omega_\uparrow$ and $\omega_\downarrow$, respectively. 
The energy level difference $\omega_0$ is $\omega_0 \equiv |\omega_\uparrow - \omega_\downarrow| = |g_n| e B_0 / 2m_p$ with $B_0$ the magnitude of $\mathbf{B}_0$. 

An axion DM background field can play the role of the oscillatory perturbation field in the Rabi oscillation of neutron spins. 
Its interaction with nucleon fields $N = (p,n)$ is described by
\begin{equation}
    \mathcal{L}_\mathrm{int}  
    = 
    - \frac{C_N}{2f_a}
    \partial_\mu a 
    \bar{N} \gamma^\mu \gamma^5 N \, ,
\label{LagNeutron}
\end{equation}
where $C_N$ is a model- and species-dependent dimensionless constant, $f_a$ is the axion decay constant, and $\bar{N}$ is the Dirac adjoint of the nucleon field. 
To be cold DM, ultralight axions can be produced non-thermally via misalignment~\cite{Preskill:1982cy, Abbott:1982af, Dine:1982ah} in the early Universe, and then form an oscillating field, $a = a_0 \cos ( m_a t - \mathbf{p}_a \cdot \mathbf{r} )$ with a mass
$m_a$ and 
momentum $\mathbf{p}_a = m_a \mathbf{v}_a$. 
The velocity $\mathbf{v}_a$ is due to the relative movement between our solar system and the Galactic halo, as well as the random velocities of the axion DM particles themselves, with a typical magnitude $v_a \equiv |\mathbf{v}_a| \simeq 10^{-3}$ in the laboratory frame. 
The amplitude $a_0$ is directly related to the 
local DM energy density $\rho_a \simeq 0.4 \, \mathrm{GeV}/\mathrm{cm}^3$~\cite{ParticleDataGroup:2024cfk} via $\rho_a = m_a^2 a_0^2 / 2$, neglecting small axion kinetic energy terms and assuming that axions make up the entirety of the DM. 
The oscillations of the axion DM field are coherent on timescales up to $\tau_\textrm{coh} \sim 2 \pi / (m_a v_a^2) \sim 2 \pi \times 10^6 / m_a$.
In the non-relativistic limit, the spatial components of the interaction in~\geqn{LagNeutron} can be simplified as~\cite{Stadnik:2013raa}, 
\begin{equation}
\label{eq:axion_wind_interaction}
    H_\mathrm{int}
    =    
    \boldsymbol{\sigma} \cdot \mathbf{B}_a \sin (m_a t) ,
    \quad
    \mathrm{with}
    \quad
    \mathbf{B}_a \equiv \frac{C_N }{2 f_a} a_0 \mathbf{p}_a \, . 
\end{equation}

Consider a beam of neutrons propagating in a uniform magnetic field with an axion DM background field acting as a perturbation.
The general form of a neutron state at time $t$ is a superposition of spin-up and spin-down states,
\begin{equation}
    \ket{\Psi(t)}
    = 
    c_\uparrow (t) e^{-i \omega_\uparrow t} \ket{\uparrow}
    +
    c_\downarrow (t) e^{-i \omega_\downarrow t} \ket{\downarrow} \, ,
\end{equation}
with two time-dependent coefficients $c_{\uparrow,\downarrow}$.
Assuming the neutrons are prepared to be spin-up initially with $c_\uparrow (0) = 1$ and $c_\downarrow (0) = 0$, these coefficients can be 
determined by solving the Schrödinger equation with the axion DM perturbation using time-dependent perturbation theory~\cite{Sakurai:2011zz}. 

In principle, the angle between the direction of the applied magnetic field ${\bf B}_0$ and the effective axion magnetic field ${\bf B}_a$ can be arbitrary, while only the component of $\mathbf{B}_a$ perpendicular to $\mathbf{B}_0$ contributes to the spin-flip of neutrons. 
To maximize the spin-flip effect, the uniform 
magnetic field $\mathbf{B}_0$ can be applied perpendicular to the direction of axion momentum, or equivalently ${\bf B}_a$, which on average is directed either along or opposite to the orbital motion of the Solar System about the Galactic Centre. 
In this case, the differential equations for the coefficients $c_{\uparrow,\downarrow}$ can be analytically solved by applying the rotating-wave approximation 
and the probability for the neutron to be in the spin-down state at time $t$ is~\cite{Sakurai:2011zz},
\begin{equation}
    P_\downarrow (\delta;t) = | c_\downarrow (t) |^2
   \simeq 
    \frac{
        |\mathbf B_a|^2 \sin ^2 
        \left( \frac{ \sqrt{|\mathbf B_a|^2+\delta^2} }{2} t \right)
    }{
        |\mathbf B_a|^2+\delta^2
    } \, ,
    \label{probability}
\end{equation}
where $\delta \equiv m_a - \omega_0$ is the difference between the axion mass-energy and the two-level energy difference. 
As a typical feature of Rabi oscillation,
this probability peaks at resonance when $m_a = \omega_0$,
\begin{equation}
    P_\downarrow^\mathrm{max} (t) 
    = 
    \sin^2 \left( \frac{1}{2} |\mathbf B_a| t \right)
    \simeq 
    \left(\frac{C_n a_0 |\mathbf p_a|}{4 f_a} t \right)^2 \, , 
    \label{max_probability}
\end{equation}
where the second equality holds for a tiny axion coupling $C_n/f_a$ or a sufficiently short time $t$. 
In deriving Eqs.~(\ref{probability}) and (\ref{max_probability}), we have treated oscillations of the axion DM field as monochromatic, which is a valid approximation for axion masses as high as $10^{-8} - 10^{-7}$\,eV for the beam parameters we consider in our present paper (see below for more details).
Near the peak at $\delta = 0$, the probability in \geqn{probability} vanishes when $\sqrt{|\mathbf B_a|^2+\delta^2} \, t / 2 = \pm \pi$. 
Thus, the minimum width of the resonance peak $\Delta \delta$ can be estimated as $\Delta \delta \simeq 4 \times 10^{-14}\,\mathrm{eV} \times (0.1 \, \mathrm{s} / t)$\footnote{The interaction time $t \simeq 0.1\, \mathrm{s}$ here corresponds to the neutron flight time $\tau$ within $\mathbf{B}_0$, discussed in more detail in the following sections.}, 
which is negligibly small for $m_a \gtrsim 10^{-12} \, \mathrm{eV}$ and the interaction times of interest in the present work. 
This narrow resonance width indicates that the Rabi oscillation approach not only aids in detecting the axion, but also enables precise measurement of its mass.

\begin{figure*}[t]
    \centering
    \includegraphics[width=\textwidth]{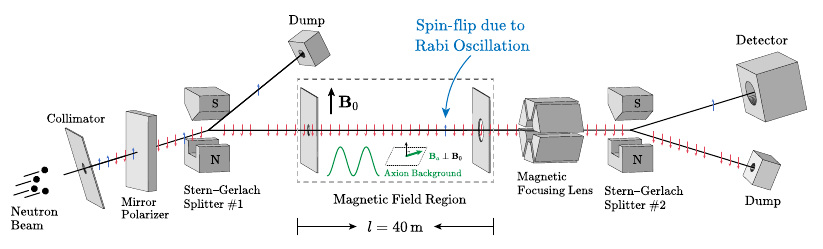}
    \caption{
    Schematic experimental setup for detecting neutron spin oscillation. 
    The incoming neutron beam is pre-polarized and then split into two beams with opposing spin states by a Stern-Gerlach apparatus. 
    The spin-down neutrons traverse a region of uniform magnetic field $\mathbf{B}_0$, where the axion DM background field induces spin-flips. 
    The spin-flipped neutrons are focused by a magnetic lens, then separated by a second Stern-Gerlach apparatus and subsequently detected.
    }%
    \label{fig:apparatus}
\end{figure*}

\textbf{Proposed Experimental Setup} -- 
The experimental setup we propose for detecting an axion DM background field via Rabi oscillation of neutron beams is illustrated in Fig.~\ref{fig:apparatus}. 
Modern advanced spallation neutron sources, such as the ESS~\cite{Garoby:2017vew}, the SNS~\cite{MASON:2006955}, and the CSNS~\cite{Wang_Sheng:2009}, can provide extremely intense neutron beams of up to about $10^{12} \,  n/\textrm{s}$. 
Initially, the neutron spins are randomly oriented.
Firstly, using a mirror polarizer, the neutron beam can be pre-polarized to a polarization state of about $99.9\%$ spin-down, losing only about $20\%$ of the neutron number in the process. Then, a Stern-Gerlach apparatus, which generates an inhomogeneous magnetic field with a vertically directed gradient, is employed to eliminate the remaining unwanted spin-up neutrons.
As neutrons traverse this magnetic-field gradient, their magnetic dipoles experience net spin-dependent forces, resulting in the splitting of spin-up and spin-down neutrons into separate beams.

The spin-down neutron beam subsequently propagates through a magnetic field region of length $l$ containing a uniform constant magnetic field directed vertically upwards. 
The coupling between neutron spins and the magnetic field creates a two-level system, with all the neutrons initially occupying the spin-down ground state. 

As discussed in the previous section, 
if the axion DM mass approaches the energy gap between the two spin states, the oscillating axion background field will induce the Rabi oscillation of neutrons in the beam due to the axion-neutron coupling. 
Consequently, a small portion of the neutrons will transition to the higher-energy spin-up state. 

After a long-distance flight, the neutron beam will spread out in the transverse directions due to velocity variations. We can add a magnetic focusing lens~\cite{Shimizu:2000aaa} to focus the beam into a smaller size. 
Finally, a second Stern-Gerlach apparatus, with the same vertically varying magnetic field, is positioned after the magnetic focusing lens to separate neutrons that have undergone a spin-flip. 
The number of spin-flipped neutrons is then counted using a neutron detector.

The main backgrounds in our experimental setup are controllable. 
Firstly, during the flight time of $\sim 0.1 \, \mathrm{s}$, approximately $10^{-4}$ 
of the neutrons decay. 
The trajectories of charged decay products ($p^+$ and $e^-$) will be strongly deflected by the Lorentz force due to the magnetic field in the second Stern-Gerlach device, 
while any remaining decay products can be easily distinguished in the neutron detector by virtue of their relatively low energies (sub-MeV versus a few MeV for the products of the reaction $n + {}^{6}\textrm{Li} \to {}^{3}\textrm{H} + {}^{4}\textrm{He}$).
Secondly, the homogeneity of the magnetic field along the flight path is at the $10^{-3}$ level. 
Neutrons may undergo spin-flips due to the magnetic-field gradient, i.e., non-adiabatic Majorana transitions. 
However, such transitions are positively correlated with the neutron velocity, as faster neutrons experience more rapid changes in the local magnetic field, whereas DM-induced spin-flips are negatively correlated with the velocity, since the DM-induced effect grows with time as shown in \geqn{max_probability}. 
This anti-correlation can help distinguish DM-induced signals from Majorana transition backgrounds. 
Finally, mHz background-rate neutron detectors based on standard techniques without spatial discrimination have been used previously, e.g., for ultra-cold neutron detection.
Neutron detectors with lower background rates of $10^{-5}$\,Hz (and below) should be feasible to construct by employing particle-physics tracking methods. The reason for this high level of discrimination is that background events emerge mainly from radioactivity in the detector materials, which can be discriminated by proper spatial resolution and fiducial cuts in the volume of interest. 
As a result, our experiment can be carried out with measurements lasting for one day at a given magnetic-field strength $B_0$ corresponding to a single specific axion DM mass. The strength of the applied magnetic field can then be incrementally varied to scan over a range of axion DM particle masses.

\textbf{Projected Sensitivity} -- 
Experimental sensitivity is enhanced with 
increased event rates. 
Neutron spallation sources can provide strong pulsed neutron beams. 
For example, at the ESS, around $10^{11}$
neutrons can be emitted per pulse within $\mathcal{O}(100)\,$ms~\cite{Santoro:2023izd}. 
Considering the $20\%$ neutron number loss in the pre-polarization stage and a factor of $4$ loss in intensity due to the divergence of the neutron beam in the transverse directions, $N_n \approx 2 \times 10^{10}$ of these are usable. 
After running the experiment for one day, $N_{\mathrm{pulse}} \simeq 10^6$ pulses can be emitted and the statistical data can reach $N_n N_{\mathrm{pulse}} \simeq 2 \times 10^{16}$. 
Typically, neutrons emitted from spallation neutron sources exhibit high velocities with a broad velocity distribution.
However, after undergoing a series of scattering and absorption processes in a decelerator, 
the neutrons can be slowed down to speeds around $v_n \sim \mathcal{O}(1000)\,$m/s,
leading to a longer interaction time with the axion DM field. 
Taking the length of the magnetic field region to be 
$l = 40\,$m as an example, the interacting time of the Rabi oscillation is 
$\tau \equiv l/v_n \sim 0.1\,$s. 
Assuming that background neutron spin-state transitions are sufficiently under control, one can estimate the projected sensitivity to the axion coupling parameters $C_n/f_a$ from the relationship that the transition probability in \geqn{max_probability} 
is less than $P_\downarrow (\tau) \lesssim 1/(N_n N_{\mathrm{pulse}})$. 
The sensitivity to the axion decay constant is optimized when the axion mass-energy matches the energy difference between different spin states in the applied magnetic field, i.e., $ \delta = 0 $, according to 
\begin{align}
\label{eq:sensitivity_estimate}
    \frac{f_a}{C_n} 
 &\gtrsim
    \sqrt{\frac{\rho_a v_a^2 \tau^2 N_n N_{\mathrm{pulse}}}{8 }}  \\ 
    &= 
    1.3 \times 10^7\,\text{GeV}
    \left( \frac{\tau}{0.1\, \mathrm{s}}\right)
    \left( \frac{N_n}{2 \times 10^{10}}\right)^{\frac{1}{2}}
    \left( \frac{N_\mathrm{pulse}}{10^6}\right)^{\frac{1}{2}} \, .  \nonumber
\end{align}
The sensitivity improves for a smaller neutron velocity and larger neutron number.
A smaller velocity results in a longer interaction time with the axion DM field, while a larger neutron number increases the overall probability of observing a spin-flip transition.

\begin{figure}[t!]
    \centering
    \includegraphics[width=8.5cm]{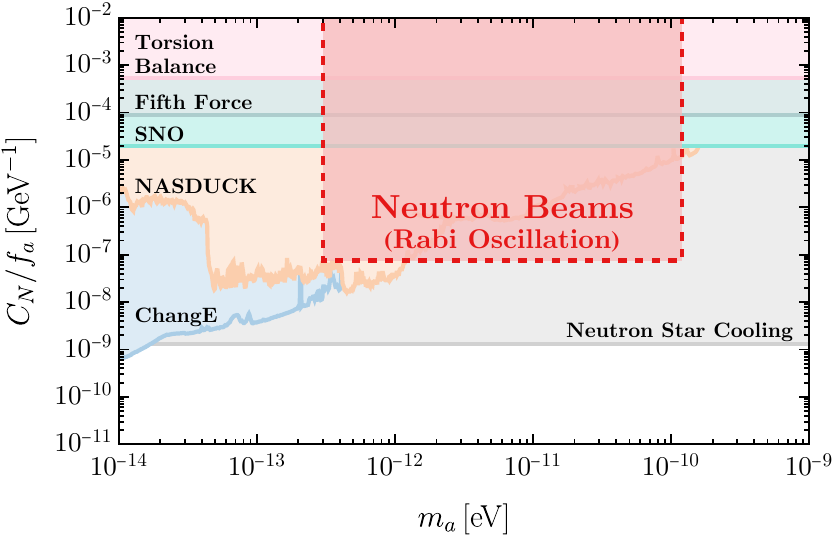}
    \caption{
    Projected sensitivity to the axion-nucleon coupling (assuming that $C_N = C_n = C_p$) assuming one day of running time at a fixed magnetic field strength $B_0$. 
    Different axion masses within the red shaded region can be probed by varying the applied magnetic field strength in the range $2.5 \times 10^{-6} - 10^{-3}$\,T. 
    The total measurement time required to scan the entire mass range is $\approx 3$ years.
    Constraints from ChangE~\cite{Wei:2023rzs,Xu:2023vfn} (light blue), NASDUCK~\cite{Bloch:2022kjm,Bloch:2021vnn} (light yellow), fifth-force experiment~\cite{Vasilakis:2008yn} (light green), torsion balance experiment~\cite{Adelberger:2006dh} (light pink), SNO~\cite{Bhusal:2020bvx} (light cyan), as well as neutron star cooling~\cite{Buschmann:2021juv} (gray), are also shown.
    }%
    \label{fig:axionN}
\end{figure}

For axion mass-energies much larger than the resonance peak width $\Delta \delta$, $m_a \gtrsim \Delta \delta$, a specific value of $B_0$ can only measure the corresponding axion mass $m_a = |g_n| e B_0 / 2m_p = 1.2 \times 10^{-10} B_0 / (10^{-3}\,\textrm{T})\,$eV. 
Since the width $\Delta \delta$ depends only on the single-shot time-of-flight $\tau$ and not on the axion mass $m_a$, lower axion masses can be scanned over much more quickly than higher axion masses. 
The magnetic field intensity that can be established stably and uniformly is in the range of $\sim 10^{-9} - 10^{-3} \, \mathrm{T}$. 
Electric dipole moment (EDM) searches using stored ultracold neutrons, such as the one in \cite{Baker:2006ts}, require volumes with magnetic-field homogeneity at the sub-$10^{-3}$ level for spin coherence. 
In the case of our proposed neutron beam measurements, we employ a similar 0.5\,m diameter and similar (albeit shorter) coil geometry. 
By incrementally varying the magnetic-field strength over the range {$2.5 \times 10^{-6} - 10^{-3}$\,T} (when the resonance is narrow and well-defined), one can achieve the projected sensitivity to axion DM in the mass range $3 \times 10^{-13} - 10^{-10}$\,eV shown as the red shaded region in \gfig{fig:axionN}. 
For our estimates, we have assumed $l=40$\,m and an integration time of 1 day at each chosen value of $B_0$ (each corresponding to a particular narrow range of axion masses). In this case, the entire indicated mass range can be scanned in $\approx 3$ years of total measurement time.

The projected sensitivity to the axion-neutron coupling in the mass range $10^{-12} - 10^{-10}$\,eV with our proposed setup can be stronger than the constraints from indirect terrestrial detection methods, such as searches for solar axions in the Sudbury Neutrino Observatory (SNO) experiment~\cite{Bhusal:2020bvx} (light cyan), the fifth-force search experiment~\cite{Vasilakis:2008yn} (light green) and the torsion balance experiment~\cite{Adelberger:2006dh} (light pink), 
as well as atomic co-magnetometry experiments, such as NASDUCK~\cite{Bloch:2022kjm,Bloch:2021vnn} (light yellow) and ChangE~\cite{Wei:2023rzs,Xu:2023vfn} (light blue). 
Besides, the axion-neutron coupling also receives strong astrophysical constraints from neutron star cooling~\cite{Buschmann:2021juv}, which is shown as the gray region in \gfig{fig:axionN}. 
However, this constraint contains some model dependencies and theoretical uncertainties.

{\bf Conclusion} -- In this paper,
we propose utilizing the 
Rabi oscillation of neutron spins in high-intensity neutron beams at spallation sources, such as the ESS, SNS and CSNS, to 
detect the axion DM background, and present a
novel experimental design to realize it.
In our setup, firstly, a Stern-Gerlach apparatus fully polarizes a 99.9\% pre-polarized neutron beam. 
Subsequently, the polarized neutrons enter a region of uniform magnetic field for free propagation. 
Due to the presence of axion dark matter, the axion-neutron coupling introduces an oscillatory perturbation to the neutron, causing the neutron spin to flip. 
Finally, a second Stern-Gerlach apparatus allows us to extract the signal of the spin-flip. 
After background and experimental sensitivity estimation and assuming the experiment runs for a day at a given applied magnetic field strength, 
our proposal can achieve a sensitivity up to two orders of magnitude stronger than the current best ground-based experimental constraints on 
the axion-neutron coupling over the mass range $10^{-12} - 10^{-10}\,$eV, reaching the level $f_a/C_n \sim 10^7$\,GeV.

By increasing the running time at a fixed magnetic field strength and the intensity of the neutron-beam flux, the sensitivity can be further enhanced. 
Once a signal is observed, our proposal also has the capability to determine the mass of the axion DM particle.

The axion DM background field can induce an oscillating neutron EDM at the one-loop level via the axion-gluon coupling~\cite{Pospelov:1999mv}.
The EDM oscillates at a frequency corresponding to the axion mass-energy along the neutron spin direction. 
If a uniform electric field $\mathbf{E}$ is applied perpendicular to the homogeneous magnetic field $\mathbf{B}_0$ inside the magnetically controlled region, the time-varying perturbation due to the coupling between the oscillating neutron EDM and the electric field will induce Rabi oscillations of the neutron spins. In order to minimise undesired systematic effects associated with the additional magnetic field $\mathbf{B}_\textrm{rel} = \mathbf{E} \times \mathbf{v} / c^2$ seen by moving neutrons with velocity $\mathbf{v}$, this electric field should be applied along the propagation direction of the neutron beam itself, which severely limits the maximum strength of the applied electric field. On the other hand, for the Ramsey approach discussed in \cite{Fierlinger:2024aik} for much lower axion masses, a strong electric field can be applied along $\mathbf{B}_0$ (which is directed transverse to the neutron beam), with undesired effects of the induced magnetic field $\mathbf{B}_\textrm{rel}$ being strongly suppressed, since the effects of the small component of $\mathbf{B}_\textrm{rel}$ directed along $\mathbf{B}_0$ on the neutrons' Larmor precession frequency are minimised in that case.

\section*{Acknowledgements}

P. F. is supported by the DFG Cluster of Excellence ``Origins'' (EXC 2094).
J. S. is supported by the National Natural Science Foundation of China (Nos. 12375101, 12425506, 12090060, 12090064) and the SJTU Double First Class start-up fund WF220442604.
The work of Y.~V.~S.~was supported by the Australian Research Council under the Discovery Early Career Researcher Award No.~DE210101593. 
C.-Y. X. is supported by the Fundamental Research Funds for the Central Universities (No. 24CX06048A).
Y.~V.~S.~gratefully acknowledges the hospitality of Tsung-Dao Lee Institute during the writing of this paper. 

\bibliographystyle{utphys}
\bibliography{ref}

\end{document}